\documentclass[longauth]{aa}  

\usepackage{subcaption}
\usepackage{graphicx}
\usepackage{xcolor}
\usepackage{txfonts}
\usepackage{hyperref}
\usepackage{xspace}
\hypersetup{
    colorlinks=true,
    linkcolor=blue,
    citecolor=blue,
    filecolor=magenta,      
    urlcolor=cyan,
    pdftitle={Overleaf Example},
    pdfpagemode=FullScreen,
    }

\makeatletter
\renewcommand*\aa@pageof{, page \thepage{} of \pageref*{LastPage}}
\makeatother

\let\oldAA\AA
\renewcommand{\AA}{\oldAA\xspace}

\newcommand{\kms}{\ensuremath{\mathrm{km\,s^{-1}}}\xspace}

\begin{document}

\title{JADES. Possible Population III signatures at z=10.6 in the halo of GN-z11}

 \titlerunning{PopIII signatures at z=10.6}
\author{
Roberto Maiolino
\inst{1,2,3}\fnmsep\thanks{rm665@cam.ac.uk}
\and
Hannah \"Ubler
\inst{1,2}
\and
Michele Perna
\inst{4}
\and
Jan Scholtz
\inst{1,2}
\and
Francesco D'Eugenio
\inst{1,2}
\and
Callum Witten
\inst{5,1}
\and
Nicolas Laporte
\inst{1,2}
\and
Joris Witstok
\inst{1,2}
\and
Stefano Carniani
\inst{6}
\and
Sandro Tacchella
\inst{1,2}
\and
William M. Baker
\inst{1,2}
\and
Santiago Arribas
\inst{4}
\and
Kimihiko Nakajima
\inst{7}
\and
Daniel J.\ Eisenstein
\inst{8}
\and
Andrew J. Bunker 
\inst{9} 
\and
Stéphane Charlot
\inst{10} 
\and
Giovanni Cresci
\inst{11} 
\and
Mirko Curti
\inst{12} 
\and
Emma Curtis-Lake
\inst{13} 
\and
Anna de Graaff
\inst{14} 
\and
Eiichi Egami
\inst{15} 
\and
Zhiyuan Ji
\inst{15} 
\and
Benjamin D. Johnson
\inst{8} 
\and
Nimisha Kumari
\inst{16} 
\and
Tobias J. Looser
\inst{1,2} 
\and
Michael Maseda
\inst{17} 
\and
Erica Nelson
\inst{18}
\and
Brant Robertson
\inst{19} 
\and
Bruno Rodríguez Del Pino
\inst{4} 
\and
Lester Sandles
\inst{1,2} 
\and
Charlotte Simmonds
\inst{1,2} 
\and
Renske Smit
\inst{20} 
\and
Fengwu Sun
\inst{15} 
\and
Giacomo Venturi
\inst{6} 
\and
Christina C. Williams
\inst{21} 
\and
Christopher N. A. Willmer
\inst{15} 
}

\institute{
Kavli Institute for Cosmology, University of Cambridge, Madingley Road, Cambridge, 
CB3 0HA, UK\\
\and
Cavendish Laboratory, University of Cambridge, 19 JJ Thomson Avenue, Cambridge CB3 0HE, UK\\
\and
Department of Physics and Astronomy, University College London, Gower Street, London WC1E 6BT, UK\\
\and
Centro de Astrobiolog\'ia (CAB), CSIC–INTA, Cra. de Ajalvir Km.~4, 28850- Torrej\'on de Ardoz, Madrid, Spain\\
\and
Institute of Astronomy, University of Cambridge, Madingley Road, Cambridge CB3 0HA, UK\\
\and
Scuola Normale Superiore, Piazza dei Cavalieri 7, I-56126 Pisa, Italy\\
\and
National Astronomical Observatory of Japan, 2-21-1 Osawa, Mitaka, Tokyo 181-8588, Japan\\
\and
Center for Astrophysics - Harvard \& Smithsonian, 60 Garden St., Cambridge MA 02138 USA\\
\and
Department of Physics, University of Oxford, Denys Wilkinson Building, Keble Road, Oxford OX1 3RH, UK\\
\and
Sorbonne Universit\'e, CNRS, UMR 7095, Institut d'Astrophysique de Paris, 98 bis bd Arago, 75014 Paris, France\\
\and
INAF - Osservatorio Astrofisco di Arcetri, largo E. Fermi 5, 50127 Firenze, Italy\\
\and
European Southern Observatory, Karl-Schwarzschild-Strasse 2, 85748 Garching, Germany\\
\and
Centre for Astrophysics Research, Department of Physics, Astronomy and Mathematics, University of Hertfordshire, Hatfield AL10 9AB, UK\\
\and
Max-Planck-Institut f\"ur Astronomie, K\"onigstuhl 17, D-69117, Heidelberg, Germany\\
\and
Steward Observatory University of Arizona 933 N. Cherry Avenue Tucson AZ 85721, USA\\
\and
AURA for European Space Agency, Space Telescope Science Institute, 3700 San Martin Drive. Baltimore, MD, 21210\\
\and
Department of Astronomy, University of Wisconsin-Madison, 475 N. Charter St., Madison, WI 53706 USA\\
\and
Department for Astrophysical and Planetary Science, University of Colorado, Boulder, CO 80309, USA\\
\and
Department of Astronomy and Astrophysics, University of California, Santa Cruz, 1156 High Street, Santa Cruz, CA 95064, USA\\
\and
Astrophysics Research Institute, Liverpool John Moores University, 146 Brownlow Hill, Liverpool L3 5RF, UK\\
\and
NSF’s National Optical-Infrared Astronomy Research Laboratory, 950 North Cherry Avenue, Tucson, AZ 85719, USA\\
}

   \authorrunning{Maiolino et al.}
   \date{}


\abstract{
Finding the first generation of stars formed out of pristine gas in the early Universe, known as Population III (PopIII) stars, is one of the most important goals of modern astrophysics. Recent models have suggested that PopIII stars may form in pockets of pristine gas in the halo of more evolved galaxies. We present NIRSpec integral field spectroscopy and micro-shutter array spectroscopic observations of the region around GN-z11, an exceptionally luminous galaxy at $z=10.6$, that reveal a greater than 5$\sigma$ detection of a feature consistent with being HeII$\lambda$1640 emission at the redshift of GN-z11.
The very high equivalent width of the putative HeII emission in this clump ($\log{(EW_{rest}(HeII)/\AA)} = 1.79^{+0.15}_{-0.25}$) and a lack of metal lines can be explained in terms of photoionisation by PopIII stars, while photoionisation by PopII stars is inconsistent with the data. 
The high equivalent width would also indicate that the putative PopIII stars likely have an initial mass function with an upper cutoff reaching at least 500~M$_\odot$. The PopIII bolometric luminosity inferred from the HeII line would be $\sim 7\times 10^{9}~L_\odot$, which 
would imply a total stellar mass formed in the burst of $\sim 2\times 10^{5}~M_\odot$. We find that
photoionisation by the active galactic nucleus (AGN) in GN-z11 cannot account for the HeII luminosity observed in the clump but can potentially be responsible for an additional HeII emission observed closer to GN-z11.
We also consider the possibility of in situ photoionisation by an accreting direct collapse black hole hosted by the HeII clump. We find that this scenario is less favoured, but it remains a possible alternative interpretation.
We also report the detection of a Ly$\alpha$ halo stemming out of  GN-z11 and extending out to $\sim$2~kpc as well as resolved funnel-shaped CIII emission likely tracing the ionisation cone of the AGN.
}

 
\maketitle

\section{Introduction} \label{sec:intro}

The formation of the first stars and galaxies marks a fundamental transitional phase in cosmic history, during which the Universe evolved from a relatively simple state into the highly structured system we observe today \citep{2004ARA&A..42...79B}. 
Almost all baryonic matter formed after the Big Bang, and the `dark ages' were composed of hydrogen and helium. Consequently, the very first stars to form must have condensed out of these elements, primarily as a result of cooling due to the small amounts of H$_2$ that could form at such early epochs \citep[e.g.][]{Kashlinsky83}.
These stars, called Population III (PopIII), are potentially very massive, up to several 100~$M_\odot$, as a consequence of inefficient gas cooling and hence poor fragmentation of pre-stellar cores at these early epochs (e.g. \citealt{Bromm99}, \citealt{Abel02},\citealt{Tan04}; \citealt{2014ApJ...781...60H}; \citealt{2014ApJ...792...32S}). 
Indirect studies of Pop III stars via stellar archaeology have confirmed these predictions, suggesting that their masses extend up to $1000~M_\odot$, with a characteristic mass greater than $1~M_\odot$ \citep{Rossi21, Pagnini23}.
Such massive PopIII stars are expected to be very hot and lead to an energetic photoionising spectrum capable of doubly ionising helium. Indeed, an expected signature of PopIII stars is prominent HeII nebular lines with very large equivalent widths (EWs; $(HeII\lambda1640)>20$~\AA) unaccompanied by metal lines (\citealt{2000ApJ...528L..65T}, \citealt{Oh01}, \citealt{2001ApJ...550L...1T}, \citealt{Schaerer03}, \citealt{nakajima_diagnostics_2022}). 
However, other models have also predicted that primordial gas might fragment more efficiently, resulting in the formation of cooler and less massive stars \citep[e.g.][]{Clark11, Greif11, Greif12, Stacy16}.

While observations have found potential chemical fingerprints of the enrichment produced by the first generation of stars (\citealt{Beers05, Frebel07, Cooke12, Hartwig15, Hartwig18, 2023arXiv230507706S, 2023ApJ...948...35S, 2023arXiv230312500K,DEugenio2023b,Christensen2023}), PopIII stars in early galaxies have eluded detection so far, with some claims not being confirmed by subsequent studies (\citealt{2015ApJ...808..139S}, \citealt{2017MNRAS.469..448B}, \citealt{2018ApJ...859...84H}). However, the unprecedented sensitivity of JWST is pushing the frontier of observations to high redshift and very faint galaxies, discovering very low metallicity systems \citep{Vanzella23} and indicating that the detection of PopIII stars may be within reach of this observatory \citep[see also e.g.][]{Zackrisson11}.

Further encouraging expectations about the detectability of PopIII stars have been obtained by recent cosmological simulations. Indeed,
two main formation phases of PopIII stars have been proposed. Firstly, PopIII stars could have formed in dark matter mini halos at redshifts z$\sim$10-30 (e.g. \citealt{1997ApJ...474....1T}; \citealt{2009Natur.459...49B}; \citealt{Pallottini14}; \citealt{2015MNRAS.452.1152J}, 
\citealt{Maio09}; \citealt{Maio10};
\citealt{Vikaeus21}; \citealt{Ventura23}; \citealt{Trinca23b}). Secondly, PopIII stars could still have formed at  later times, down to redshifts of about z $\sim $ 3-6, in pristine or low-metallicity pockets of gas due to the highly inhomogeneous nature of metal enrichment in the Universe, as well as a consequence of re-accretion of pristine gas (e.g. \citealt{2007MNRAS.382..945T}; \citealt{2010MNRAS.404.1425J}; \citealt{2020MNRAS.497.2839L}, \citealt{2018ApJ...854...75S}; \citealt{2019MNRAS.488.2202J}, \citealt{2022arXiv220704751K}, \citealt{Volonteri23}).
 In these cases, PopIII stars are expected to be found in the halos of more massive galaxies, and their hard ionising spectrum may be ionising mildly enriched gas, resulting in weak emission of metal lines (\citealt{nakajima_diagnostics_2022}, \citealt{2022arXiv220704751K}, \citealt{Jaaks17}).

 Although PopIII stars are expected to be present even at relatively low redshifts (possibly down to z$\sim$3 ; \citealt{2013A&A...556A..68C}, \citealt{2020MNRAS.497.2839L}) in the latter scenario, the probability for a halo to host PopIII stars is higher at very high redshift \citep{trinca+2023}. 
Within this context, GN-z11 is an exceptional galaxy at z=10.6, as it offers an optimal environment to search for PopIII stars. 
Identified initially by \cite{Bouwens10} and \cite{Oesch16},
GN-z11 is the most luminous galaxy at z$>$10 in all Hubble Space Telescope (HST) fields (including all fields in the Cosmic Assembly Near-infrared Deep Extragalactic Legacy Survey, CANDELS, and all Frontier Fields, \citealt{2018ApJ...855..105O}). Recent JWST-NIRCam imaging from the JWST Advanced Deep Extragalactic Survey \citep[JADES][]{Eisenstein2023} has revealed that GN-z11 is characterised by an unresolved central source, a more extended component (200 pc in radius) with S\'ersic index n=1, and a more diffuse component extending for $\sim$0.5$''$ to the NE (the `haze'), which may not be associated with GN-z11 and could simply be a foreground system \citep{tacchella_jades_2023}. \cite{bunker_jades_2023} obtained low- and medium-resolution spectroscopy of GN-z11 with JWST-NIRSpec in the multi-object spectroscopy (MOS) mode using the micro-shutter assembly (MSA), revealing a spectrum with a wealth of nebular emission lines in various ionisation stages and characterised by some unusual line ratios. Interestingly, the spectrum revealed Ly$\alpha$ emission extended along the slit for about $0.2''$ ($\sim 0.8$~kpc) towards the SW (the position angle of the shutter was 19.9$^\circ$). Additional NIRSpec MSA observations targeting GN-z11 have led to the identification of an active galactic nucleus (AGN) in GN-z11 \citep{Maiolino2024Natur}, primarily through the detection of high ionisation lines (NeIV), semi-forbidden lines tracing very high densities consistent with the broad-line regions of AGNs, and the detection of the CII$^*\lambda$1335 fluorescent emission, typically seen only in AGN. The spectrum of GN-z11 has also revealed the presence of a powerful outflow traced by the redshifted ($\sim 530$ \kms) and resonantly scattered emission of both Ly$\alpha$ and CIV$\lambda$1550 and the blueshifted absorption of CIV ($\sim 800-1000$ \kms).
The second NIRSpec MSA observation had a position angle nearly perpendicular to the first observation and did not show obvious evidence for Ly$\alpha$ extension on either side of GN-z11.

A more detailed analysis of the 2D spectrum of GN-z11 in the first observation has revealed the tentative emission of a line located at $\sim$0.5$''$ to the NE of GN-z11 at a wavelength consistent with HeII$\lambda$1640 at the same redshift of GN-z11 (see Sect. \ref{ssec:heii}). This intriguing finding prompted a DDT proposal to obtain NIRSpec integral field unit (IFU) spectroscopy of GN-z11 with both the G140M (3.3~h) and G235M (10.6~h) gratings.
The IFU datasets deliver superior spatial information relative to the MSA (although the sensitivity is lower than the MSA because the IFU 
 has an about 50\% lower throughput).

In this paper we report a preliminary analysis of the new NIRSpec-IFU data of GN-z11 that confirm the HeII$\lambda$1640 detection at z=10.6 as well as additional interesting features associated with the Ly$\alpha$ and CIII] morphology. Throughout this work, we use the AB magnitude system and assume a Planck flat $\Lambda$CDM cosmology with $\Omega_m=0.315$ and $H_0=67.4$ km/s/Mpc \citep{Planck20}.  With this cosmology, $1''$ corresponds to a transverse distance of 4.08 proper kpc at $z=10.6$.

\begin{figure}[h]%
\centering
\includegraphics[width=1.0\columnwidth]{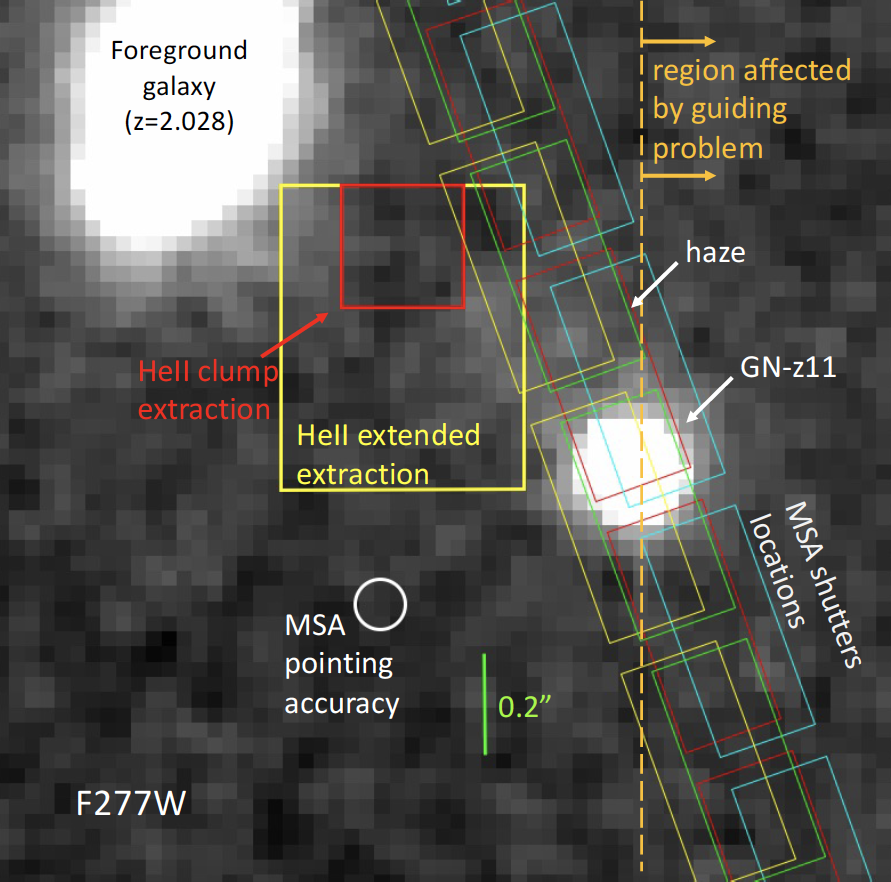}
\caption{
F277W NIRCam image of the central $\sim 1.7''$ around GN-z11 illustrating the location of the MSA shutters during the first NIRSpec-MOS observations in February 2023 (tilted boxes; boxes of the same colour are associated with the same configuration), the extraction region of the HeII clump (red square), and the HeII extended emission (yellow box). The white circle indicates the absolute uncertainty of the MSA target acquisition (hence the shutters' footprints are uncertain by this amount). North is up, and east is to the left. The orange vertical dashed line marks the boundary beyond which fewer dithers from the NIRSpec-IFU observations are available because of the guiding problem reported in Sect.~\ref{sec:data}. We note that the full IFU FoV is larger than shown here ($>$3$''\times$ 3$''$ taking into account the dithers).
}\label{fig:footprints}
\end{figure}

\begin{figure*}[h]%
\centering
\includegraphics[width=2.0\columnwidth]{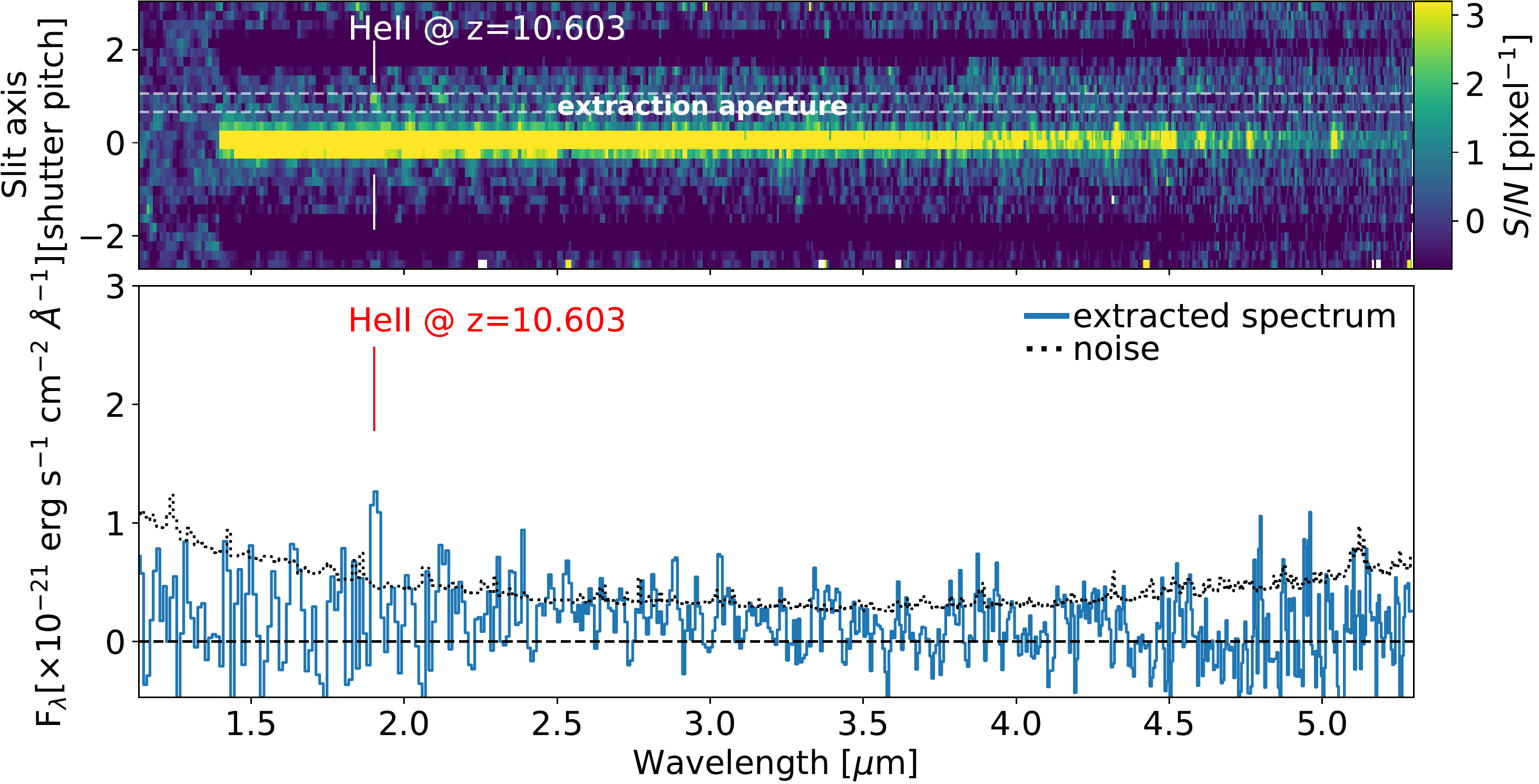}
\caption{
Prism spectrum of the region to the north-east of GN-z11, from the February 2023 MSA observation.
Top panel: Signal-to-noise 2D prism spectrum of GN-z11 . The spectrum has been obtained by using only the exposures in which GN-z11 is in the upper and lower shutters (hence discarding the exposure with GN-z11 in the central shutter) as to avoid self-subtraction of the extended signal.
Bottom panel: One-dimensional spectrum extracted from the region delimited by the two dashed horizontal lines, at about 0.45$''$ from GN-z11 in the NE direction,
illustrating the tentative detection (3.7 $\sigma$ integrated) of a feature at 1.903 $\mu$m.
The red vertical line indicates the location of HeII$\lambda$1640 expected at the redshift of GN-z11, z=10.603 (which is indistinguishable from the redshift of 10.600 inferred for the grating spectrum).
}\label{fig:prism}
\end{figure*}

\section{Observations, data processing, and data analysis}\label{sec:data}

GN-z11 was observed with NIRSpec-IFU \citep{Jakobsen22,boker22} on 22 to 23 May 2023 under the DDT programme 4426 (PI: Roberto Maiolino). The observational setup utilised a medium cycling pattern of 10 dithers for a total integration time on source of 10.6~h with the medium-resolution grating-filter pair G235M/F170LP, and 3.3~h with the medium-resolution grating-filter pair G140M/F100LP. These configurations cover the spectral ranges of 0.97–1.89 and 1.66–3.17  $\mu$m, with a nominal spectral resolution of $R\sim1000$.  The detector was set with the improved reference sampling and subtraction pattern (IRS$^2$), which significantly reduces the read out noise with respect to the conventional method \citep{Rauscher12}.  We also checked that no bright foreground sources were at the location of the MSA quadrants, where they could significantly contaminate the IFS spectra. 

Although JWST has an excellent pointing accuracy (i.e. 0.1$''$, 1~$\sigma$ radial\footnote{\url{https://jwst-docs.stsci.edu/jwst-observatory-characteristics/jwst-pointing-performance}} \citealt{Rigby23}), and the telescope  was commanded to the location of GN-z11 provided by the NIRCam and HST imaging and GAIA-referenced astrometry, the observations were affected by a guiding problem
of the observatory because the guide star turned out to be a binary (report from the Fine Guidance Sensor team). This introduced a large offset of about 1.4$''$ towards the SE.
This offset located GN-z11 as being close to the edge of the IFU field of view (FoV). This restricted the zone around GN-z11 that could be explored, lowering the S/N of regions that were covered by fewer dither positions or excluding completely some regions from the FoV. More specifically, out of the 10 dither positions, GN-z11 itself was located right at the edge of the FoV in three of them, and in another four of them GN-z11 was so close to the edge that the SW extension of the Ly$\alpha$ was mostly outside the FoV. As we attempt an analysis of the SW extension based on the incomplete data in this paper, our work on that regard will need to be validated by planned future observations that will repeat the dithers that were more severely affected (within the context of the same programme 4426). As a consequence, the main focus of this paper is on the extended emission towards the NE.

Because of the guiding problem, the world coordinates given by the pipeline in the header of the final IFU cube are off by about 1.4$''$. Therefore, we corrected the world coordinates by realigning the cube to the NIRCam image by collapsing the cube in the same band as in the NIRCam filter (although the centroid of the continuum does not change with wavelength).

Raw data files were downloaded from the MAST archive and subsequently processed with the JWST Science Calibration pipeline\footnote{\url{https://jwst-pipeline.readthedocs.io/en/stable/jwst/introduction.html}} version 1.8.2 under CRDS context jwst\_1068.pmap. We made several modifications to the default reduction steps to increase the data quality, which are described in detail by \cite{Perna23} and which we briefly summarise here. 
The default pipeline is occasionally subject to over-subtraction of elongated cosmic ray artefacts through the default circular correction applied to `snowball-like' cosmic ray hits. To avoid this problem, we patched the pipeline to fit ellipses to all flagged regions consisting of five or more adjacent pixels. The regions with best-fit ellipses with axial ratios smaller than 0.1 were removed from the list of `snowball-like' cosmic rays. 
Count rate frames were corrected for $1/f$ noise through a polynomial fit.
Outliers were flagged on the individual 2D exposures using an edge-detection algorithm similar to {\sc lacosmic} \citep{vDokkum01}. We calculated the derivative of the count rate maps along the x-axis of the detector (on the scale of a few pixels, the x-axis is a good approximation for the dispersion direction). The derivative was normalised by the local flux (or by three times the rms noise, whichever was highest), and we rejected the 95\textsuperscript{th} percentile of the resulting distribution --
the percentiles were calculated over the entire detector area \citep[see ][ for details]{Deugenio2023a}.
In addition, we made the following corrections to the *cal.fits files after Stage 2: We masked pixels at the edge of the slices (two pixels wide) to conservatively exclude pixels with unreliable sflat corrections. We also masked regions affected by leakage from failed open MSA shutters. We note that the data quality flag `MSA\_FAILED\_OPEN' fails to account for all failed open MSA shutters. At the same time, several regions marked as being affected by failed open shutters do not show any impact from leakage, and we included those regions to increase $S/N$ in the final combined cube. Finally, we masked regions that were affected by persistence in all dither frames (e.g.\ from cosmic rays). 
The final cubes were combined using the `drizzle' method with pixel scales of $0.06^{\prime\prime}$ and $0.10^{\prime\prime}$, for which we used an official patch to correct for a known bug.\footnote{\url{https://github.com/spacetelescope/jwst/pull/7306}}

\begin{figure*}[h]%
\centering
\includegraphics[width=2.0\columnwidth]{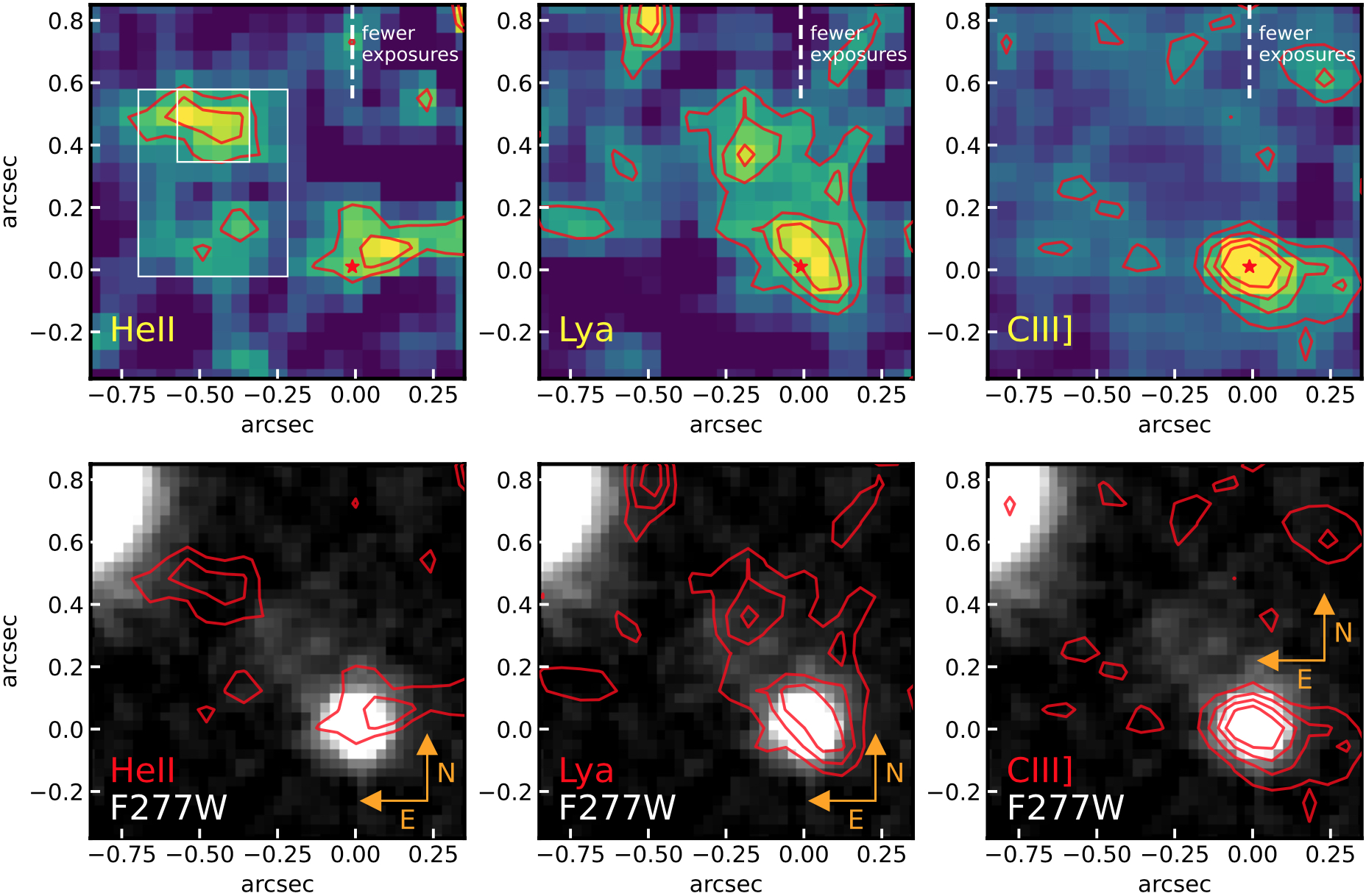}
\caption{
Emission lines maps. Top panels: Maps of the HeII$\lambda$1640, Ly$\alpha$, and CIII]$\lambda$1909 lines (see text for details). The red star marks the position of the continuum of GN-z11. Contours are at 2, 3, and 4 $\sigma$ for the HeII and Ly$\alpha$ maps and at 2, 4, 6, and 8 $\sigma$ for the CIII] map (the rms is estimated from the statistics in the IFU FoV, outside this region). We note that most of these spaxels are independent; hence the significance of the integrated emission of each feature is significantly higher. The two white boxes indicate the extraction apertures of the two spectra shown in Fig.~\ref{fig:spectra_ex} and marked in red and yellow in Fig.~\ref{fig:footprints}. The vertical dashed segment indicates the  region beyond which there are fewer exposures because of the guiding problem discussed in Sect. \ref{sec:data}. Bottom panels: Overlay of the contours from the emission line maps onto the F277W NIRCam map from \cite{tacchella_jades_2023}.
}\label{fig:maps}
\end{figure*}

We also re-analysed the NIRSpec-MOS observations of GN-z11 obtained in February 2023 within the JADES survey but with a focus on the 2D extraction along the shutters. These observations were described in \cite{bunker_jades_2023}. Here, we only remind that those observations were obtained with four configurations
of the MSA \citep{Jakobsen22,Ferruit22,boker23}, each with three shutters nodding. The low-resolution prism and the three medium-resolution gratings were used, for a total observing time of 6.9 hours with the prism and 3.45 hours with each of the gratings. In the following, to explore the 2D information we use only the prism, as the gratings exposures are shallower.
The processing of the MSA data is also described in \cite{bunker_jades_2023}; here, we only recall that we have used the pipeline 
developed by the ESA NIRSpec Science Operations Team and the NIRSpec GTO Team. We also recall that the spatial pixel size in the MSA 2D spectra is 0.1$''$ and that each shutter is 0.2$''$ wide.
The standard background subtraction method does not allow for exploration of the extended emission along the shutters. Indeed, the nodding technique moves the target by one shutter position (i.e. by 0.$''$53) along the shutter's direction. When reciprocally subtracting two exposures in which the source is located in adjacent shutters, this results in a deep negative trace of the subtracted source spectrum at $\pm$0.5$''$ from the (positive) spectrum of the source and at least as wide as the PSF and its wings ($\sim$0.2$''$), depending on the source size. This makes the standard 2D spectrum unusable in the exploration of any extended emission beyond $\sim$0.2$''$ from the source. Therefore,
in order to explore the 2D information along the shutter and avoid self-subtraction of the extended emission, we used a different background subtraction process relative to standard process adopted in \cite{bunker_jades_2023}. Specifically, we did not use the exposure with the source in the central shutter. Instead, we used only the exposures in which the source is in the upper and lower shutters (i.e. the exposures with the source in the most distant positions) to perform mutual background subtraction. This ensured that the subtracted negative traces of the GN-z11 spectrum are located at $\sim 1''$ from the (positive) spectrum and enabled the study of spectral features along the shutter up to about 0.6$''$ from the source. This method obviously slightly reduces the sensitivity of the data, as we only used two-thirds of the data, and the noise in the background was increased by a factor of $\sqrt{2}$, but it is far superior relative to the standard method for exploring extended emission.

The tilted boxes in Fig.\ref{fig:footprints} show the footprint of the shutters in the four MSA configurations of the MOS observation from February 2023 overlaid on the F277W image of GN-z11 \citep{tacchella_jades_2023}. The white circle indicates the target acquisition uncertainty of the MSA, meaning that the footprints indicated in the figure may actually be displaced in any direction by that amount. The  orange vertical dashed line indicates the region beyond which fewer exposures from the NIRSpec-IFU DDT programme are available (and with GN-z11 at the edge of the field of view) because of the telescope guiding problem discussed above.

When extracting spectra and emission line maps, we removed background and continuum emission with the following procedure: We extracted the background from an area of the IFU free from artefacts and performed sigma-clipping of the outliers. The resulting spectrum was smoothed with a rolling median of 25 spectral pixels. This background was normalised to the continuum in the vicinity of the spectral feature and subtracted. Since the background has some residual structure across the IFU FoV, also because of the possible continuum contribution from individual sources (especially in the case of GN-z11), we further optimised the continuum subtraction by linearly interpolating the continuum within a range of a few 0.01 $\mu$m around the line of interest after masking the emission line (typically within $\pm$ 3 spectral pixels from the line peak).
In the case of the Ly$\alpha$, the latter step could be more problematic on GN-z11, as the galaxy continuum is asymmetric relative to Ly$\alpha$ due to the intergalactic medium (IGM) absorption (which also affects the continuum red-wards of the Ly$\alpha$ because of the damping wing). However, we were primarily interested in the extended emission where the galaxy continuum is not detected. However, we also explored the subtraction of the continuum around Ly$\alpha$  by fitting the continuum only on the red side of the line, and the result did not change, except for a slight increase in the noise on the extended component, as expected from the poorer sampling of the background.

\begin{figure*}[h]%
\centering
\includegraphics[width=1.7\columnwidth]{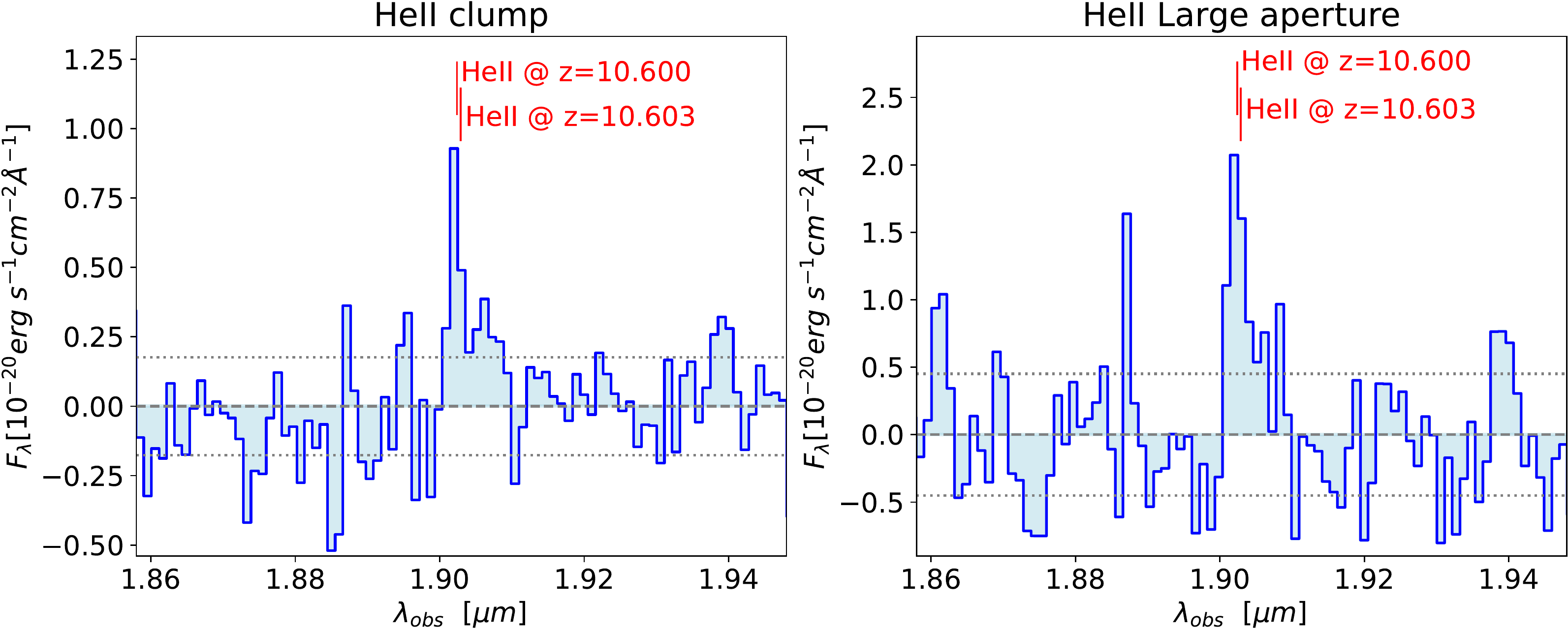}
\caption{
Spectra extracted from the apertures indicated in Fig. \ref{fig:footprints} and zoomed near the expected wavelength of HeII at the redshift of GN-z11. Specifically, the left panel shows the spectrum extracted from the aperture centred on the `HeII clump' (red box in Fig.\ref{fig:footprints} and smaller white box in Fig.\ref{fig:maps}). The right panel shows the spectrum extracted from the larger aperture encompassing the extended HeII emission in the NE quadrant of GN-z11 (yellow box in Fig.\ref{fig:footprints} and larger white box in Fig.\ref{fig:maps}). The red vertical bars indicate the expected wavelength of HeII at the redshift of GN-z11, z=10.603, and the wavelength of the centroid of the line in the clump, which, if identified as HeII, gives a redshift z=10.600. The dotted lines indicate the noise per spectral pixel in each spectrum inferred from the rms directly measured in the spectrum around the HeII line.
}\label{fig:spectra_ex}
\end{figure*}

\section{Results}\label{sec:results}

\subsection{HeII detection in the halo of GN-z11}\label{ssec:heii}

We start by illustrating the tentative detection of HeII$\lambda$1640 in the 2D spectrum along the shutters in the NIRSpec MSA observation of February 2023. We show the S/N 2D spectrum of GN-z11 in the top panel of Fig.~\ref{fig:prism}. As already discussed, this was reprocessed to use only the exposures in which GN-z11 is in the upper and lower shutters (hence discarding the exposure with GN-z11 in the central shutter) in order to prevent self-subtraction of extended emission.
The 2D spectrum shows a tentative signal at 1.905 $\mu$m, which is consistent with the expected location of HeII$\lambda$1640 at the redshift of GN-z11 (z=10.603), peaking at about 0.5$''$ from GN-z11.
The bottom panel shows the spectrum extracted from the region of the 2D spectrum delimited by the two horizontal dashed lines in the upper panel. The black dotted line indicates the noise level per spectral pixel. The extracted spectrum confirms the emission feature at 1.905 $\mu$m. The integrated significance of the feature is 3.7$\sigma$, as reported in Table~\ref{tab:emlines}. 
We note that this feature is totally impossible to detect with the standard background subtraction procedure adopted in \cite{bunker_jades_2023}, because in that case the deep negative trace of the GN-z11 subtracted spectrum falls exactly at this location. This is the first negative trace, next to the (positive) spectrum of GN-z11, in the top panel of Fig.1 of \cite{bunker_jades_2023}. With our method, we avoided the negative traces nearest the GN-z11 spectrum (i.e. those at $\pm$0.5'' from the spectrum of GN-z11 in the 2D spectrum of \citet{bunker_jades_2023}).
We also note that the absolute flux and spatial centroid of the feature are affected by some additional uncertainty associated with the 2D extraction. Specifically, the feature is at the edge of the region beyond which self-subtraction can still happen, even when adopting our two-shutter methodology. This can result in both flux uncertainty and the centroid being slightly artificially shifted towards GN-z11. Secondly, in some exposures, the feature falls next to or beneath the shutter bar (see Fig. \ref{fig:footprints}). This can also affect the flux and centroid. Finally, the aperture losses applied to the spectrum are based on the point-like assumption and on using the location of GN-z11 (which is next to the bar or to the slit edge in some exposures). Hence, these corrections may result in additional flux uncertainties when applied to the putative HeII feature, which is located in another region of the shutters.

Figure \ref{fig:maps} shows the maps at the expected wavelengths of HeII, Ly$\alpha$, and CIII] obtained from the NIRSpec-IFU observation. Specifically, HeII was extracted from 1.9003$\mu$m to 1.9035$\mu$m, CIII] from 2.2099$\mu$m to 2.2173$\mu$m, and Ly$\alpha$ from 1.4115$\mu$m to 1.4152$\mu$m. The extraction of CIII] and Ly$\alpha$ is on a slightly broader range, as the former is a doublet (hence, it is slightly broader, although still unresolved at our resolution), and Ly$\alpha$ has a broad profile. We note that the Ly$\alpha$ map has been centred on the Ly$\alpha$ offset observed in GN-z11 ($-550$ \kms). The maps have also been slightly smoothed with a Gaussian kernel with a width of 1.5 pixels (0.09$''$) .

Since the emission features are extended, the significance of the integrated emission of the individual clumps is higher than that given by the contours.
The HeII map shows a plume extending for about 0.25$''$ towards the west of GN-z11 (red star).
This is the area with fewer exposures (due to the guiding problem), so it should be treated with great care. Yet, if it is confirmed, such a HeII plume would extend well beyond the size of the galaxy identified by \cite{tacchella_jades_2023} and may be tracing gas photoionised by the AGN. The weak CIII] at this location (and the fact that no CIV emission is seen either) would indicate that this is very low metallicity gas photoionised by the AGN.

The most intriguing feature in Fig. \ref{fig:maps} is the clump located at about 0.6$''$ (2.4~kpc) to the NE. This is very close to the location of the emission serendipitously found along the shutter of the MSA prism observation (Fig. \ref{fig:prism}), especially when also taking into account the MSA target acquisition uncertainty and the fact that the MSA observation still suffers signal self-subtraction at locations farther than 0.5$''$ from GN-z11 (as well the additional uncertainties discussed in the previous section). We extracted the spectrum from this region by taking a $0.24''\times 0.24''$ square aperture (red box in Fig.~\ref{fig:footprints}; `HeII clump extraction', smaller white box in Fig.~\ref{fig:maps}; and Tab.~\ref{tab:apertures}). The resulting spectrum is shown in the top-left panel  of Fig. \ref{fig:spectra_ex}, which shows the detection of a line at $\lambda _{obs}=1.902~\mu$m whose integrated significance is 5.5~$\sigma$ (Table \ref{tab:emlines}). This corresponds to the wavelength of HeII$\lambda$1640 at a redshift of 10.600, fully consistent with the redshift of GN-z11. 

It is unlikely that this feature is H$\alpha$  or [OIII]$\lambda$5007 of a lower redshift interloper, specifically at z=1.90 or z=2.80,  as other bright lines expected at these redshifts are not seen. The `haze' identified in imaging by \cite{tacchella_jades_2023} is probably a lower redshift interloper, but it is offset (closer to GN-z11) relative to the putative HeII clump. We indeed identified an emission line at 2.330 $\mu$m peaking on the `haze' that we tentatively identify as [OIII] at z=3.65 (see Appendix \ref{app:lowz_int} for details), consistent with the photometric redshift inferred by \cite{tacchella_jades_2023}. 
The large foreground galaxy at z=2.028 to the NE of the clump does not have any line emission at this wavelength (see spectrum in Appendix \ref{app:lowz_int}), and any weak emission feature potentially associated with that galaxy would also show a much stronger H$\alpha$ associated with it, as well as many other stronger emission lines.

We also explored whether some Ly$\alpha$ emission is leaking out of the intergalactic medium (IGM) absorption at this location. We found a tentative detection of Ly$\alpha$, but it requires additional data for confirmation. Therefore, we do not discuss this any further.

It is difficult to compare the fluxes of the putative HeII seen in the IFU clump and the detection in the MSA. In terms of wavelength, they are fully consistent. In terms of flux, it is more difficult. At face value, the fluxes reported in Tab.\ref{tab:emlines} give a higher flux in the MSA (with a large error). However, because of the uncertainty on the exact location of the MSA shutter (see Sect.\ref{sec:data} and white circle in Fig.\ref{fig:footprints}), the actual fraction of the putative HeII clump that entered the shutters is not known. Moreover, the potential absolute flux calibration issues plaguing the MSA at the location of the putative HeII mession, as discussed above, introduce additional uncertainties that prevent a proper comparison. Finally, we note that the MSA prism observation is more sensitive than the IFU observation (as discussed in Sect.\ref{sec:intro}). Hence, it may be detecting additional HeII emission outside the clump not seen in the IFU cube. Additionally, faint emission at slightly different velocities (as discussed in Appendix \ref{app:lowz_int}) may be missed by the higher resolution of the grating, but picked by the much broader line spread function of the prism.

In addition to the clump at $\sim$0.6$''$ to the NE from GN-z11, the HeII emission appears to have additional and more extended emission, as seen in the map in Fig. \ref{fig:maps} (top left), mostly distributed in the NE quadrant of GN-z11, possibly with a fainter and less significant clump about 0.3$''$ south of the HeII primary clump and about 0.4$''$ east of GN-z11. We extracted the spectrum from this larger area by using the large aperture marked with the yellow box in Fig. \ref{fig:footprints} (larger white box in Fig.\ref{fig:maps}). The resulting spectrum at the wavelength of HeII at z=10.600 is shown in the right panel of Fig. \ref{fig:spectra_ex}. In this case, the putative HeII emission is detected at 6.1$\sigma$ and with a total flux that is about two times higher than in the clump. We note that the putative HeII emission detected in this larger aperture is certainly not polluted by HeII emission in GN-z11. To begin with, the HeII emission of GN-z11 is very faint \citep{Maiolino2024Natur,bunker_jades_2023}; contamination by GN-z11 would imply that all other strong lines seen in GN-z11 would be detected and would be much brighter than HeII. Secondly, as shown in Appendix \ref{app:sp_ap_mod}, we also extracted a spectrum by modifying this large aperture in order to remove the region closer than 0.3$''$ (more than six times the PSF radius at this wavelength) from GN-z11, and the detection remained unchanged (Fig.\ref{fig:spheiimod}), therefore confirming that the weak HeII emission from GN-z11 does not contribute at all to the spectrum extracted in this region.

\subsection{\texorpdfstring{Ly$\alpha$}{Lya} halo and morphology}\label{ssec:lya}

Although not the focus of this paper, in the middle panel of Fig. \ref{fig:maps}, we also show the map of the Ly$\alpha$ emission. Here, we have simply collapsed the background- and continuum-subtracted channels around the redshifted peak observed in GN-z11 (\ref{ssec:heii}). A more in-depth analysis of the full Ly$\alpha$ profile will be presented in a forthcoming paper. The Ly$\alpha$ emission is clearly extended, partly in the western direction relative to GN-z11 (red star) but mostly towards the SW, as expected from the MOS data when taken along this orientation \citep{bunker_jades_2023}. The extension towards the SW is very sharp and could potentially trace an accreting filament. However, as discussed in the next section, the fact that we also see lopsided CIII] emission in this direction favours the interpretation that the emission in the SW stems from gas illuminated in the ionisation cone of the AGN in GN-z11. We, however, warn that this region falls in the area with less exposure because of the telescope guiding problem discussed in Sect. \ref{sec:data}, so the emission in this region could potentially extend farther to the SW. A proper investigation  requires additional data that will be obtained in the repeated observations.

At a low surface brightness level, Ly$\alpha$ extends smoothly and over a larger area also towards the north, with a peak at about 0.4$''$ to the NE of GN-z11.
The nature of the extended Ly$\alpha$ emission will be discussed more extensively in a dedicated paper.

\subsection{CIII] extension and the ionisation cone}\label{ssec:ciii}

The CIII]1909 doublet is one of the brightest metal lines observed in the spectrum of GN-z11, and it is the only metal line of GN-z11 that is clearly resolved in the IFU observations. Its continuum-subtracted 
map is shown in the right panel of Fig. \ref{fig:maps}. The CIII] emission is clearly elongated in the NE-SW direction and most prominently extended towards the SW, with a funnel-shaped geometry. We note that the elongation is not an artefact of the data processing, as we demonstrate in the Appendix (where we compare it with the continuum map extracted around the same wavelength). The most plausible interpretation is that this emission is tracing the ionisation cone of the AGN in GN-z11. Although no CIV was clearly detected in this region \citep[which however might be self-absorbed;][]{Maiolino2024Natur}, the emission line properties are fully consistent with an AGN photoionisation scenario with a low ionisation parameter, likely due to the distance from the AGN \citep{Feltre16,nakajima_diagnostics_2022}. Once again, since this extended emission falls in the region where the exposure time is shorter, because of the telescope guiding  problem discussed above, the investigation of this extended emission therefore requires the additional data that will be obtained in the repeated observations.

\begin{figure}[h]%
\centering
\includegraphics[width=1.0\columnwidth]{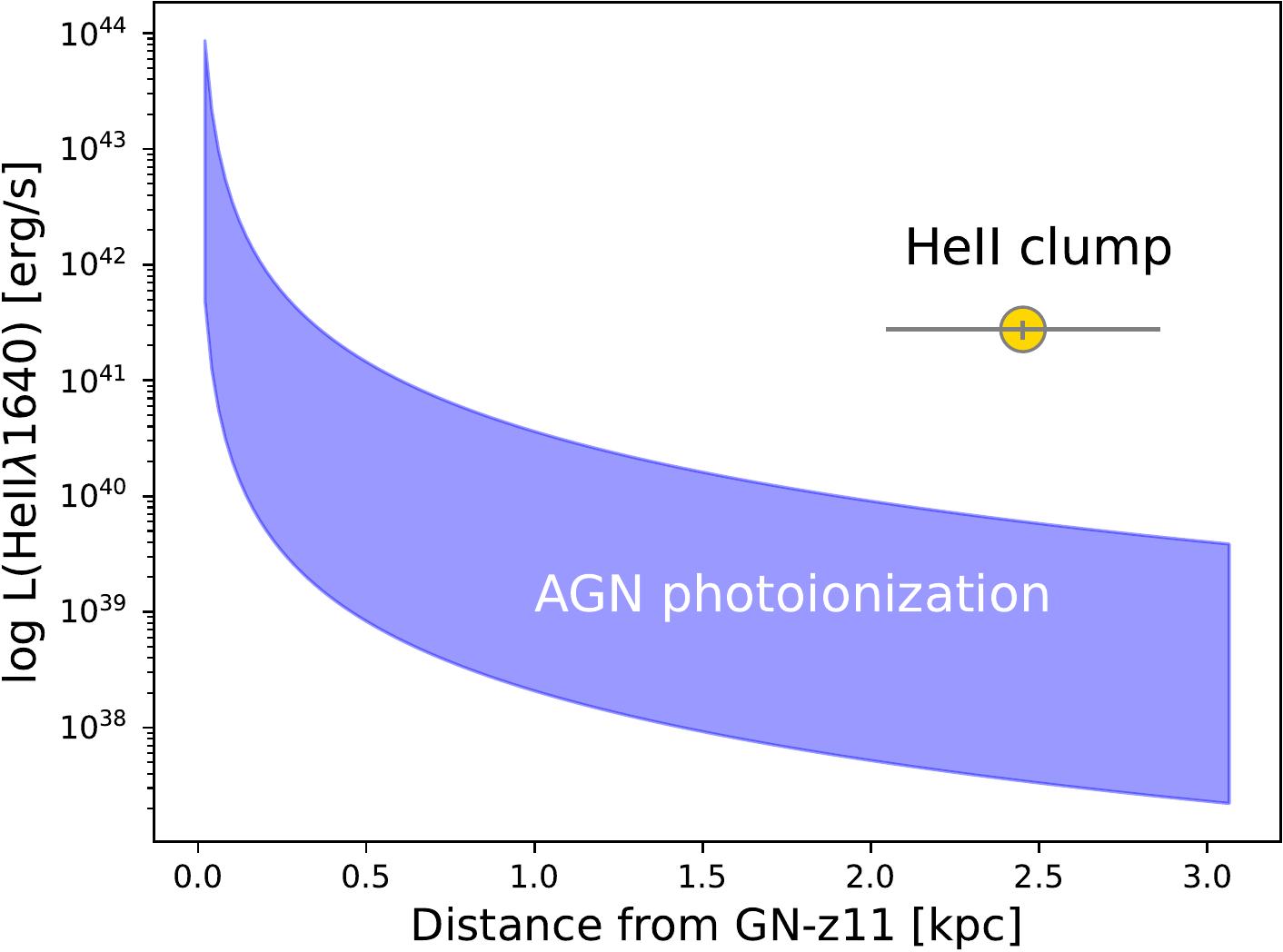}
\caption{
Observed HeII$\lambda$1640 luminosity of the `HeII clump' and its projected distance from GN-z11 compared to the expected HeII luminosity as a function of distance from GN-z11, in case of AGN photoionisation. In this calculation, we used the upper limit for the size of the HeII clump and assumed a broad range of AGN ionising shapes powering the photoionisation. Further conservative assumptions are that the `HeII clump' is not made of unresolved clumps and that the projected distance is the actual separation. Clearly, the HeII luminosity observed in the `HeII clump' is inconsistent with being photoionised by the AGN in GN-z11.
}\label{fig:lheii}
\end{figure}

\begin{figure*}[h]%
\centering
\includegraphics[width=1.9\columnwidth]{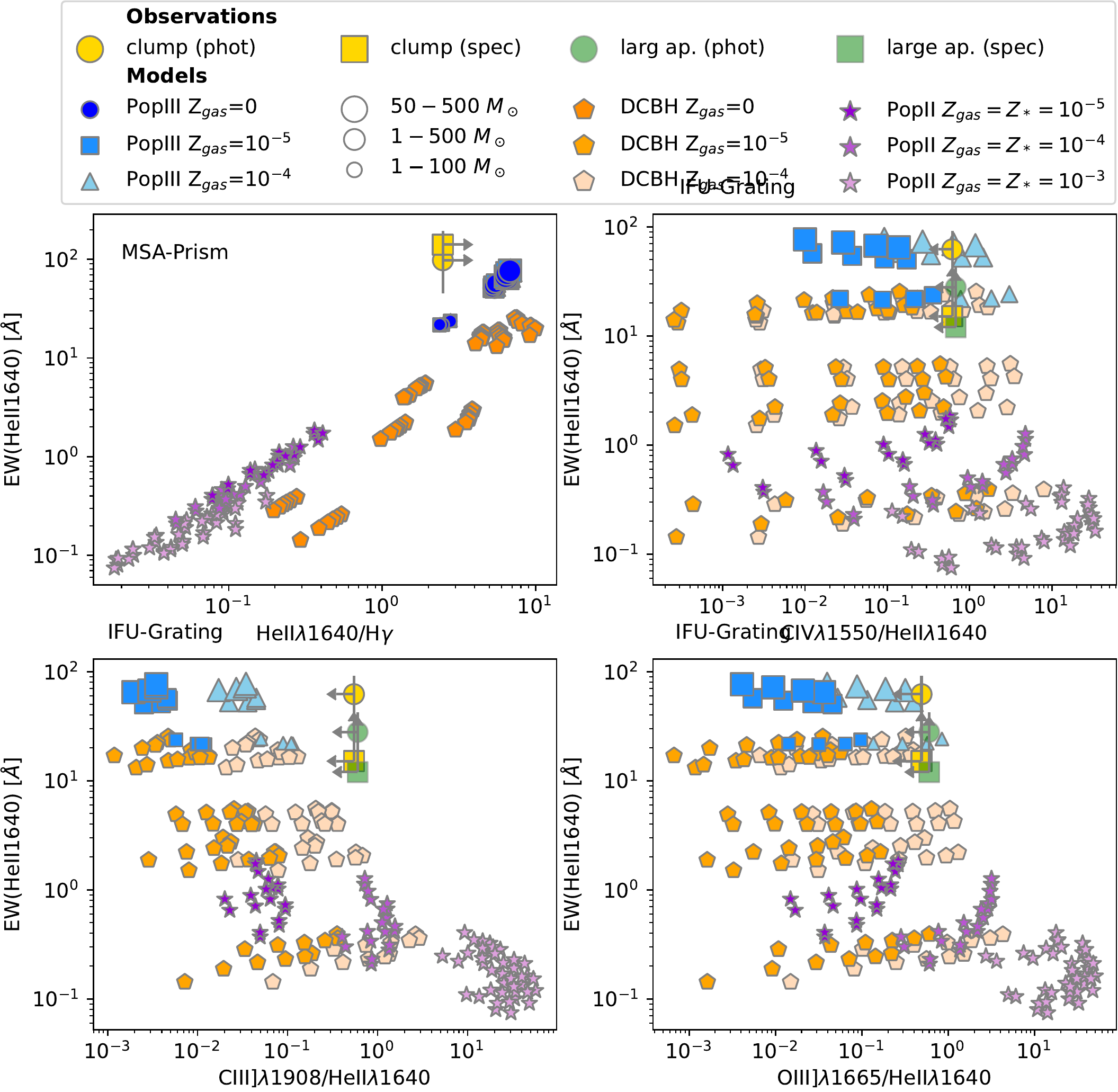}
\caption{
Diagnostic diagrams showing the EW of HeII$\lambda$1640 versus various line ratios: HeII/H$\gamma$, CIV$\lambda$1550/HeII, CIII]1909/HeII, and OIII]$\lambda$1665/HeII.
The observational constraints on the HeII clump in the halo of GN-z11 are shown with golden symbols, while the observational constraints for the HeII emission obtained from the large aperture extraction are shown with green symbols. Circles are for EWs for which the continuum has been inferred from photometry, while squares show the EWs for which the continuum constraints have been inferred directly from the spectra.
Purple stars show PopII models with different (absolute) metallicities (colour-coded with different purple shadings) and where the gas and stars are assumed to have the same metallicity.
Blue circles are models of PopIII stars in a pristine gas; light blue squares are models of PopIII stars in a mildly enriched gas, specifically $\rm Z_{gas} = 10^{-5}$;
light blue triangles are models of PopIII stars in a more enriched gas, specifically $\rm Z_{gas} = 10^{-4}$. The PopIII markers have different sizes depending on the stellar mass range assumed for the IMF, as indicated in the legend. Orange pentagons show the predictions for accreting DCBHs embedded in gas with different (absolute) metallicities. 
}\label{fig:heii_d}
\end{figure*}

\section{The case for pristine gas and Population III}\label{sec:popiii}

As discussed in Sect. \ref{ssec:heii}, the most likely identification of the emission line observed at 1.902~$\mu$m in the halo of GN-z11 is the HeII$\lambda$1640 emission at z=10.600. In the vicinity of GN-z11, specifically the western plume and some of the extended emission towards the NE quadrant, the HeII emission may likely trace gas photoionised by the AGN in GN-z11. The fact that no or little emission from metal lines (in particular CIV, CIII], and OIII]) was detected in these regions suggests that such circumgalactic gas has a low metallicity. However, it is also possible that the low intensity of all metal lines is a consequence of the low ionisation parameter  \citep{Feltre16,nakajima_diagnostics_2022}.
Yet, empirically, 
when extended HeII emission is observed in other halos surrounding AGNs at lower redshift, 
this is always accompanied by even brighter CIV emission and possibly CIII] emission \citep{Guo20,Fossati21}. Therefore, the finding of extended HeII emission around an AGN at z$>$10 not accompanied by CIV emission suggests that this is primarily due to much lower gas metallicity in the halo.

Although the gas closer to GN-z11 may be photoionised by the AGN, the HeII emission seen in the NE clump at $\sim$0.6$''$ (2.4 kpc) from GN-z11 cannot be explained in terms of photoionisation by the AGN simply based on ionising photon budget arguments. Indeed, HeII$\lambda$1640 is a simple recombination line that is essentially a counter of photons more energetic than 54 eV from the impinging  radiation. As a consequence, by knowing the luminosity of the AGN  \citep{Maiolino2024Natur} as well as the distance and size of the cloud (or its upper limit), one can estimate the number of HeI ionising photons reaching the cloud and the resulting HeII luminosity. We assumed a broad range of AGN ionising spectra \citep{nakajima_diagnostics_2022} and that the observed projected distance gives the actual distance of the clump to GN-z11. This obviously is a conservative assumption, as the actual distance is larger and results in a lower flux of ionising photons. We also assumed that within the clump, the covering factor of the absorbing gas is unity. In other words, we have made the additional conservative assumption that the gas in the cloud is not `clumpy', and hence all radiation reaching it is fully absorbed.
The result of this calculation is shown in Fig. \ref{fig:lheii}, where the blue shaded area shows the expected HeII luminosity as a function of distance from GN-z11, assuming photoionisation by the AGN in GN-z11 and by adopting all AGN ionising continuum shapes considered in \cite{nakajima_diagnostics_2022}. We also assumed isotropic emission, which is another conservative assumption given that the AGN in GN-z11 is type 1 \citep{Maiolino2024Natur}. The luminosity observed in the HeII clump ($\rm L(HeII) = 2.8\times 10^{41}~erg/s$, based on the flux reported in Table \ref{tab:emlines})\footnote{We note that no correction for extinction was attempted, as we did not see any evidence for dust extinction at this location (and not even in GN-z11). Nonetheless, this is a conservative approach, as any dust extinction would make the HeII luminosity even higher.} is shown with the golden symbol, which is clearly inconsistent with AGN photoionisation by more than one order of magnitude, even ignoring that all assumptions made are very conservative. We note that we have assumed stable emission, but it is very unlikely that the luminosity of the AGN can change by two orders of magnitude because the variability observed so far is constrained to the level of 10\% \citep{Maiolino2024Natur}.

The HeII emission in the clump must therefore result from in situ photoionisation. In Fig. \ref{fig:heii_d}, we explore the source of ionising photons and gas metallicity by using diagnostic diagrams in which the EW of the HeII line is 
compared with various line ratios, specifically: HeII/H$\gamma$, CIV$\lambda$1550/HeII, CIII]$\lambda$1909/HeII, and OIII]$\lambda$1665/HeII.
We compared these quantities with the photoionisation models presented in \cite{nakajima_diagnostics_2022}. Specifically, the purple stars jn Fig. \ref{fig:heii_d} illustrate the case of photoionisation by PopII stars (through metal-poor models from the BPASS libraries) with a broad range of ionisation parameters and densities and with (absolute) metallicities ranging from 10$^{-5}$ to 10$^{-3}$. In these models, the gas and stars are assumed to have the same metallicity. The blue symbols show the models for PopIII ionising continua. The dark blue circles are the case of PopIII embedded in pristine gas. The light blue squares and cyan triangles show the case of PopIII photoionising slightly enriched gas, as indicated in the legend. In the case of PopIII, \cite{nakajima_diagnostics_2022} used a Salpeter IMF with three different stellar ranges: 1--100 M$_{\odot}$, 1--500 M$_{\odot}$, 
and 50--500 M$_{\odot}$. In Fig. \ref{fig:heii_d}, these ranges are coded with PopIII symbols of different sizes, as indicated in the legend. We note that despite the high mass range explored in \cite{nakajima_diagnostics_2022}, a Salpeter IMF may not be appropriate for PopIII \citep[e.g.][]{Bennassuti14, Bennassuti17, Stacy16, Venditti23}, and the EW(HeII) for PopIII may therefore have values even higher than those obtained by those models.

The constraints on the HeII clump in the halo of GN-z11 are shown with golden symbols.
The continuum is totally undetected in the IFS medium-resolution spectra and only marginally detected in the low-resolution prism MSA spectrum. Therefore, we primarily used the method adopted by various other studies of determining the continuum estimating the EW from the imaging photometry \citep[e.g.][]{Stark17,Debarros17,Mainali17,Jung20,Kusakabe20,Boyett22,Matthee23a}.
The constraints on the HeII clump using the continuum determination from photometry are shown with golden circles in Fig.\ref{fig:heii_d}. We note that the top-left diagram is only for the MSA prism measurement, as we needed a constraint on H$\gamma$, while the other panels are for the IFU+grating measurements. 
We also provide (golden squares) the constraints obtained from the continuum non-detection in the IFS medium-resolution spectroscopic data as well as the estimation of the EW of the clump directly from the MSA low-resolution spectrum. A more detailed description of the derivation of the EW constraints is given in Appendix \ref{app:ew}.
The resulting EWs are given in Table \ref{tab:emlines}. The EW(HeII) inferred for the IFS spectrum of the clump and for the MSA spectrum are consistent with each other given the uncertainties.

For the HeII/H$\gamma$ ratio, we used the data from the prism, as H$\gamma$ is not covered by the new IFU observations. (We note that the HeII/H$\gamma$ is a lower limit, given that H$\gamma$ was not detected; see Table \ref{tab:emlines}.) In this case, we measured the EW consistently from the MSA prism spectrum, both by extracting the continuum from photometry at the corresponding aperture along the MSA shutter (golden circle) and by directly estimating the continuum from the spectrum (golden square), as reported in Table~\ref{tab:apertures}. 

We also report the constraints on the HeII emission extracted from the larger aperture (green symbols). Although, as mentioned, in this case, part of the emission may be associated with the AGN photoionisation.

Figure~\ref{fig:heii_d} illustrates that the high EW(HeII$\lambda1640)$ observed in the HeII clump, specifically $\log{(EW(HeII)/\AA)} = 1.79^{+0.15}_{-0.25}$ from the IFU and $\log{(EW(HeII)/\AA)} = 1.99^{+0.18}_{-0.33}$ from the MSA,  is inconsistent by nearly two orders of magnitude with the PopII models but consistent with the PopIII models. Furthermore, it requires an IMF with a very large upper mass cutoff, about 500~M$_\odot$. To explain the very large EW(HeII), an even higher stellar mass cutoff is possibly needed or an IMF more top-heavy than Salpeter in addition to the high mass cutoff or perhaps a log-flat IMF \citep{Chon21} is required.

We emphasise that in order to be consistent with the PopII models, the continuum flux of the clump should be about 80~nJy (i.e. nearly as strong as GN-z11), which is totally inconsistent with the data and therefore excludes the PopII scenario with a very high level of confidence. To make our point more clearly, we note that the 2$\sigma$ lower limit on the EW(HeII) from the photometry method is 20\AA, which is totally inconsistent with the PopII scenario.

The lower limit on the HeII/H$\gamma$ ratio (Fig.~\ref{fig:heii_d}, top left; see also Table~\ref{tab:emlines}) is  consistent with the constraint from the EW(HeII), that is, requiring photoionisation from PopIII stars and possibly having an IMF with a very massive upper mass cutoff. 
In contrast, the upper limits from the metal lines (see Table~\ref{tab:emlines}) are not very constraining in terms of metallicity, and we cannot exclude that the PopIII stars are embedded in some moderately enriched gas, although the CIV/HeII upper limit is at the borderline with the models of PopIII living in a gas with a metallicity of 10$^{-4}$.

By using the PopIII models with the massive upper cutoff IMFs (1-500~$M_\odot$ and 50-500~$M_\odot$), we inferred the ratio between HeII luminosity and total bolometric luminosity ($\rm \sim 10^{-2}$ for the two  IMF models with the highest cutoff masses, which are better matches with the observations). This gave a bolometric luminosity of $\rm 2.7\times 10^{43}~erg~s^{-1}$ for the PopIII in the clump. Assuming an Eddington limited luminosity, this gave a mass of only $\rm 2\times 10^5~M_\odot$. 
Alternatively, by using the ZAMS mass to luminosity conversions for PopIII given by \cite{Schaerer02}, the inferred bolometric luminosity would give a total stellar mass of $\rm 2.5\times 10^5~M_\odot$, similar to the value obtained above assuming an Eddington limited luminosity.
These values for the total PopIII stellar mass are not far from the PopIII stellar mass expected from some recent simulations. 
Indeed, these simulations expect a halo hosting a PopII stellar mass similar to GN-z11 ($\rm \sim8\times 10^8~M_\odot$ in the extended component, \citealp{tacchella_jades_2023}) to harbour a fractional mass of PopIII stars of about $\rm 3\times 10^{-4}-1\times 10^{-3}$, that is, a total PopIII stellar mass of $\rm \sim2-8\times 10^{5}~M_\odot$ (\citealt{Yajima23, Venditti23}; A.~Venditti, priv.\ comm.).

It is worth noting that in addition to the extended HeII emission and although the map is noisy, the HeII clump itself seems resolved and possibly extended over $\sim$0.25$''$ (i.e. $\sim$1 kpc). This is much larger than the expected size of a single PopIII sub-halo, and it is much more likely that the extension is associated with a cluster of PopIII sub-halos. Indeed, according to recent simulations, PopIII sub-halos are expected to cluster in regions of a few kiloparsecs \citep{Tornatore07, Venditti23}. Alternatively, the extension may trace a large clump of pristine gas that is being photoionised by a single halo of PopIII stars embedded in it.

Finally, we also considered the possibility that the HeII clump is ionised in situ by an accreting direct collapse black hole (DCBH) seed in a pristine, or very low-metallicity, gas cloud. \cite{nakajima_diagnostics_2022} have explored this case in their photoionisation calculations, and their models are shown with orange pentagons in Fig.~\ref{fig:heii_d}.
The darker orange symbols show the case of a DCBH photoionising pristine (zero metallicity) gas, while lighter shades of orange indicate the case of moderately enriched gas, as indicated in the legend. As for the PopII and PopIII cases shown in the same figure, the DCBH models explore a broad range of densities and ionisation parameters. Additionally, the models explore a broad range of plausible ionising shapes for the accreting DCBH \citep[we defer to ][for a more detailed description of the models]{nakajima_diagnostics_2022}. The DCBH models reach HeII EWs that are much larger than PopII stars; however, they are not as high as the PopIII stars with the very massive upper cutoff IMFs. In particular, the DCBH models do not reach the EW(HeII) required to explain the observation of the HeII clump. However, the difference is not large (a factor of about two to four within the error bars), and we did not exclude that more extreme DCBH models could be consistent with the observations. Therefore, although the DCBH scenario is less favoured, it remains a possible alternative interpretation.

\section{Conclusions}

We have presented new NIRSpec-IFU observations of GN-z11, a remarkably luminous galaxy at z=10.6. We have also presented a reanalysis of the 2D MSA spectra of GN-z11 from the JADES survey by employing a data processing technique that enables the analysis of extended emission along the shutters. Unfortunately, a telescope guiding problem during the IFU observations resulted in GN-z11 being offset to the western edge of the FoV in various dither positions, so the focus of the paper has been on the analysis of the extension in the eastern part of the FoV, while the properties of the western extension were only briefly analysed (a more detailed analysis of the western extension is deferred until after repeated observations of the affected dithers, within the context of the same observing programme).
The main observational findings are the following:

\begin{itemize}

\item We detected a spectral feature at 1.902 $\mu$m in a clump at $\sim 0.5-0.6''$ ($\sim$ 2~kpc) to the NE of GN-z11. This wavelength corresponds to HeII$\lambda$1640 at z=10.600, and it is fully consistent with the redshift of GN-z11. The line was detected both in the MSA prism (3.7$\sigma$) and in the IFU grating (5.5$\sigma$).

\item The HeII emission was also detected over a more extended area in the NE quadrant of GN-z11, possibly with a second (fainter) clump. There is also a HeII plume extending towards the west of GN-z11.

\item Ly$\alpha$ and CIII] were also clearly extended and well resolved. Ly$\alpha$ has a sharp feature extending towards the SW of GN-z11, where bright CIII] also extends in a funnel-shaped geometry (however, we remind that these extended features towards the west will need to be re-examined with new data).
At a lower surface brightness, Ly$\alpha$ also extends over a larger region of up to $\sim$2~kpc to the NE of GN-z11 and has a peak at about 1.5~kpc from the galaxy.
\end{itemize}

We investigated these features in detail, mostly focusing on the HeII emission, and reached the following conclusions:

\begin{itemize}

\item While the HeII emission close to GN-z11 may be associated with photoionisation by the AGN, simple photon budget arguments exclude that this is the case for the HeII clump, leaving in situ ionisation as the only possibility.

\item We showed that the very high EW of HeII ($\log{(EW(HeII)/\AA)} = 1.79^{+0.15}_{-0.25}$ in the IFS and $\log{(EW(HeII)/\AA)} = 1.45^{+0.18}_{-0.32}$ in the MSA) and the large HeII/H$\gamma$ ratio cannot be explained in terms of photoionisation by PopII (metal-poor) stars, but they are consistent with photoionisation by PopIII stars.

\item The very high EW(HeII) suggests that the putative PopIII stars must have an IMF reaching an upper mass cutoff of at least 500~M$_\odot$. The non-detection of H$\gamma$ (hence the lower limit on HeII/H$\gamma$) also supports this scenario.

\item We infer that the mass of PopIII stars formed in the burst is about $\rm \sim 2\times 10^5~M_\odot$, which is not far from the fraction of PopIII stars in massive halos at these redshifts expected by some simulations.

\item We also considered the alternative possibility of photoionisation by a DCBH in the HeII clump. This scenario is less favoured, as it predicts a lower EW, but not by a large factor. Hence, this scenario remains another possible interpretation.

\item Finally, we suggest that the funnel-shaped CIII] extension, accompanied by the Ly$\alpha$ elongated feature towards the SW, traces the ionisation cone of the AGN hosted in GN-z11.

\end{itemize}

These results have demonstrate the JWST's capability to explore the primitive environment around galaxies in the early Universe, revealing fascinating properties. Additional observations, especially with NIRSpec-IFS, will provide additional crucial information, especially for differentiating some of the scenarios discussed above.

\begin{table*}
    \centering
    \caption{Extraction apertures. Given the IFU observation guiding problem, the IFU mask coordinates were applied only once the cubes were registered to have the GN-z11 continuum matching the absolute coordinates given in \cite{tacchella_jades_2023}.}
    \begin{tabular}{lcccc}
    \hline
   & RA & DEC & size($''$) & PA \\
    \hline
HeII clump & 12:36:25.514 & 
+62:14:31.80 &
0.24$\times$0.24 &
0 \\
HeII large ap. & 12:36:25.514 &
+62:14:31.62 &
0.48$\times$0.60 &
0 \\
MSA offset ap. & 12:36:25.471 &
+62:14:31.78 &
0.20$\times$0.20 & 19.9 \\
 \hline            
    \end{tabular}
    \label{tab:apertures}
\end{table*}

\begin{table*}
    \centering
    \caption{Measured fluxes, EWs, and $3\sigma$ upper limits in the apertures considered in this paper.}
    \begin{tabular}{lccc}
    \hline
    Emission line & Flux  &  log(EW)[phot] & log(EW)[spec]\\
                   &  $10^{-19}$ erg s$^{-1}$ cm$^{-2}$  & \AA & \AA \\
    \hline

    \multicolumn{4}{c}{HeII clump small aperture} \\
     HeII$\lambda$1640      & 1.8$\pm$0.34 &
        $1.79^{+0.15}_{-0.25}$ & $> 1.18$\\
   CIV1550 & $<$ 1.2 & & \\
   OIII]1665 & $<$ 1.0 & & \\
   CIII]1909 & $<$ 1.1 & & \\
    \hline
    \multicolumn{4}{c}{HeII large aperture} \\
     HeII$\lambda$1640      & 5.0$\pm$0.83 & $1.45^{+0.18}_{-0.32}$ & $> 1.08$\\
   CIV1550 & $<$ 3.1 & & \\
   OIII]1665 & $<$ 2.5 & & \\
   CIII]1909 & $<$ 2.8 & & \\
    
    \hline

    \multicolumn{4}{c}{MSA offset aperture} \\
     
     HeII$\lambda$1640      & 3.4$\pm$0.9 &
     $1.99^{+0.18}_{-0.33}$ & $2.16^{+0.13}_{-0.21}$\\
    H$\gamma$ & $<$ 1.37 & & \\
    \hline
    
    \end{tabular}
    \label{tab:emlines}
\end{table*}

\begin{table*}
    \centering
    \caption{Continuum flux densities in HeII clump from NIRCam.}
    \begin{tabular}{lc}
    \hline
    \hline
    Filter & Flux [nJy]   \\
    \hline
        \multicolumn{2}{c}{HeII clump small aperture} \\
     F150W      & 4.4$\pm$1.8 \\
     F200W      & 2.2$\pm$1.6 \\
     F277W      & 3.1$\pm$0.9 \\
    \hline
       \multicolumn{2}{c}{HeII large aperture} \\
     F150W      & 19.7$\pm$6.2 \\
     F200W      & 23.0$\pm$7.8 \\
     F277W      & 25.9$\pm$8.1 \\
 \hline
         \multicolumn{2}{c}{MSA offset aperture} \\
     F150W      & 2.2$\pm$1.8 \\
     F200W      & 3.6$\pm$1.6 \\
     F277W      & 4.9$\pm$0.9 \\
 \hline            
    \end{tabular}
    \label{tab:phot}
\end{table*}

\begin{acknowledgements}

We thank Alessandra Venditti, Harley Katz, Raffaella Schneider, Richard Ellis, and Stefania Salvadori for valuable comments.
FDE, JS, RM, TL, WB acknowledge support by the Science and Technology Facilities Council (STFC), ERC Advanced Grant 695671 ``QUENCH" and by the UKRI Frontier Research grant
RISEandFALL. RM also acknowledges funding from a research professorship from the Royal Society.
AJB acknowledges funding from the ``FirstGalaxies" Advanced Grant from the European Research Council (ERC) under the European Union’s Horizon 2020 research and innovation programme (Grant agreement No. 789056).
BER, BJ, CNAW, DJE, EE, FS acknowledge support from the NIRCam Science Team contract to the University of Arizona, NAS5-02015. DJE is also supported as a Simons Investigator.
The research of CCW is supported by NOIRLab, which is managed by the Association of Universities for Research in Astronomy (AURA) under a cooperative agreement with the National Science Foundation.
CW thanks the Science and Technology Facilities Council (STFC) for a PhD studentship, funded by UKRI grant 2602262.
ECL acknowledges support of an STFC Webb Fellowship (ST/W001438/1).
GC acknowledges the support of the INAF Large Grant 2022 ``The metal circle: a new sharp view of the baryon cycle up to Cosmic Dawn with the latest generation IFU facilities".
GV, SCa acknowledge support by European Union’s HE ERC Starting Grant No. 101040227 - WINGS.
H{\"U} gratefully acknowledges support by the Isaac Newton Trust and by the Kavli Foundation through a Newton-Kavli Junior Fellowship.
JW acknowledges support from the ERC Advanced Grant 695671, ``QUENCH'', and the Fondation MERAC.
NL acknowledges support from the Kavli foundation.
PRdP, MP, SA acknowledge support from Grant PID2021-127718NB-I00 funded by the Spanish Ministry of Science and Innovation/State Agency of Research (MICIN/AEI/ 10.13039/501100011033). MP also acknowledges support from the Programa Atraccion de Talento de la Comunidad de Madrid via grant 2018-T2/TIC-11715.
RS acknowledges support from a STFC Ernest Rutherford Fellowship (ST/S004831/1).  MM acknowledges support from the National Science Foundation via AAG 2205519.

\end{acknowledgements}

\bibliographystyle{aa}
\bibliography{roberto2}

\appendix

\section{NIRSpec-IFU contiuum map}\label{app:cont_map}

\begin{figure}[h]%
\centering
\includegraphics[width=0.95\columnwidth]{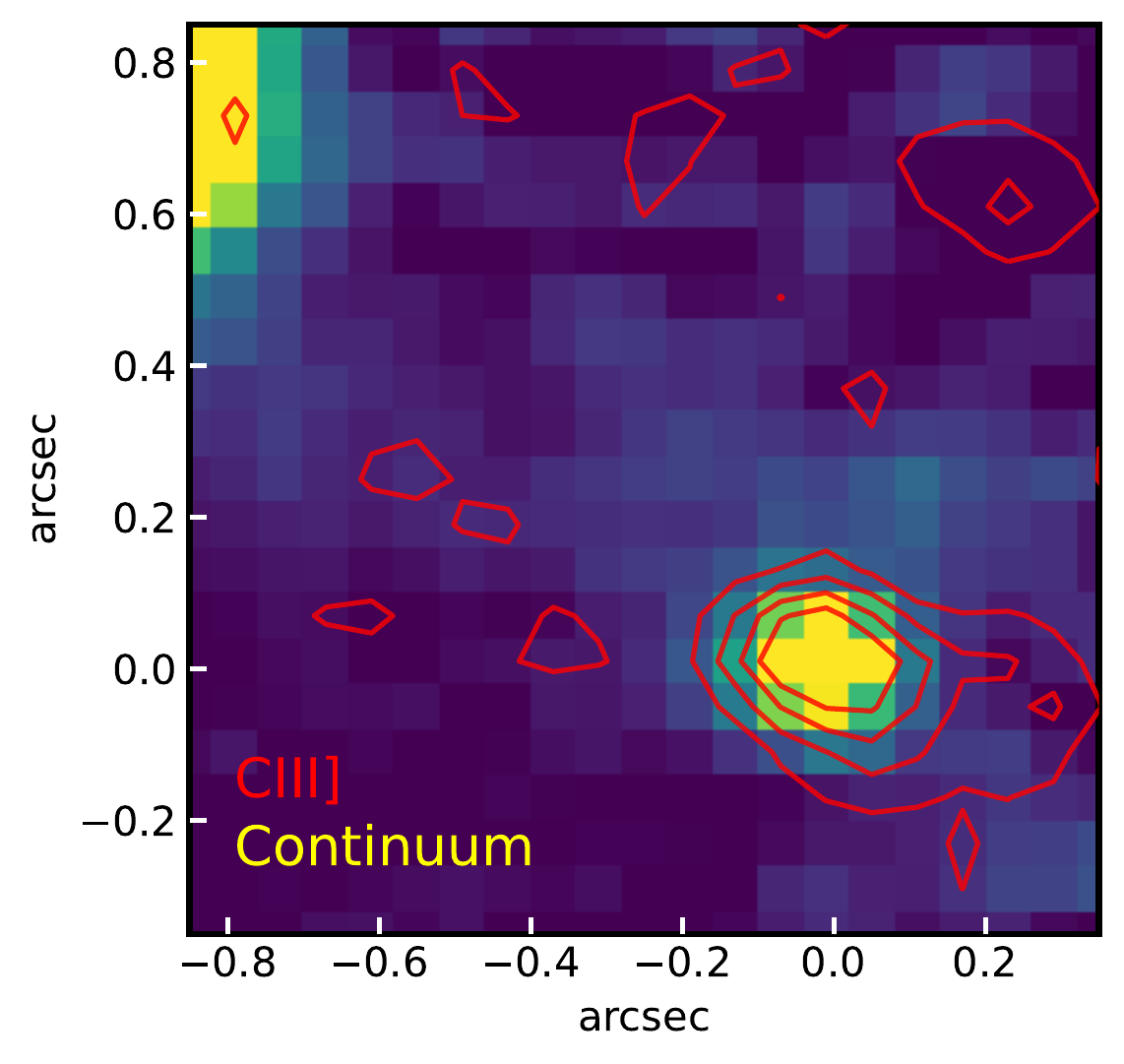}
\caption{Map of the GN-z11 continuum around the CIII] emission line with the CIII]  contours overlaid (red). Contours are 2, 4, 6, and 8 $\sigma$ levels. The CIII] emission shows an extension to the East, while the continuum appears like a point source.
}\label{fig:IFS_cont_CIII}
\end{figure}

Figure.\ref{fig:IFS_cont_CIII} shows the map of the continuum obtained by collapsing the continuum channels near the CIII]1909 emission with contours of the continuum-subtracted CIII] line map overlaid. The continuum map does not show any elongation, demonstrating that the CIII] elongation is not an artefact of the data processing.

\section{Interlopers and other emission-line sources}\label{app:lowz_int}

\begin{figure*}%
\centering
\includegraphics[width=0.95\textwidth]{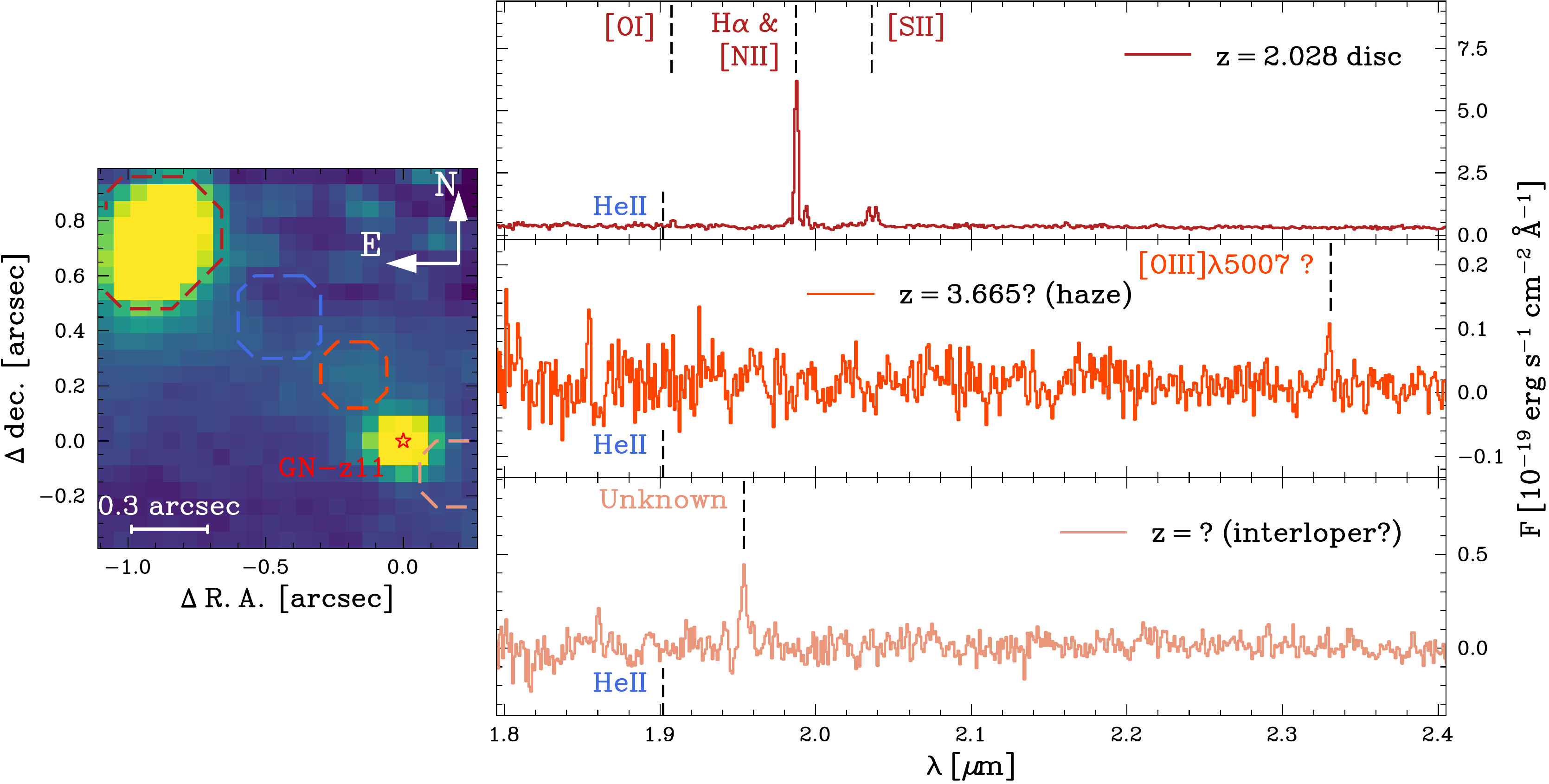}
\caption{Map of the continuum in a region centred at the `haze' highlighting a bright
disc galaxy at $z=2$, the haze itself, and a region south-west of GN-z11. The spectra
of these three regions are shown in the right panels, highlighting the wavelength
of our HeII detection, as well as other detected emission lines. The galaxy at $z=2$ does
not have any bright feature at wavelengths close to 1.902~$\mu$m. The haze presents
a single emission line that we tentatively identify as [OIII]$\lambda$5007 at $z=3.66$
-- which would be consistent with the photometric redshift from \citet{tacchella_jades_2023}.
Finally, we detected an emission line in the SW lobe of GN-z11. At the redshift of GN-z11,
this line would correspond to a rest-frame wavelength of 0.1684~$\mu$m -- which we do not
trace to any notable emission line.
}\label{fig:interlopers}
\end{figure*}

We searched the region around GN-z11 for possible interlopers. There is a disc galaxy with
bright H$\alpha$ emission 1.1~arcsec north-east of GN-z11. We integrated the flux of this
galaxy (dark red contours in Fig.~\ref{fig:interlopers}, left) and show the resulting spectrum
(top-right panel). At the redshift of $z=2.028$, there is no strong emission line that may
contaminate the HeII emission (i.e. near $\lambda = 1.90\;{\rm \mu m}$). Besides, at the
location of the HeII `clump' (blue contours), the strength of any emission line due to the
interloper would be even weaker than what is shown in the extracted spectrum. Conversely, if the putative HeII emission in the clump were due to some faint line associated with the galaxy at z=2.028, then the spectrum extracted from the same aperture of the HeII clump would also show much stronger emission of H$\alpha$ and the several other strong emission lines, which are not seen.
We can therefore exclude that the line we identify as HeII is due to a misidentified faint line
emission at $z=2.028$.

The red contours show the extraction box around the `haze' \citetext{cf. \citealp{tacchella_jades_2023}}.
The corresponding spectrum shows an emission line at $\lambda = 2.331 \; {\rm \mu m}$. We tentatively
identify this line as either [OIII]$\lambda$5007 -- typically the strongest emission line in low-mass
galaxies. This would mean the `haze' was a $z=3.665$ interloper -- consistent with the photometric
redshift of \citet{tacchella_jades_2023}. If the line was H$\alpha$, we would expect to see
[OIII]$\lambda$5007 at a bluer wavelength (low-mass galaxies typically have little dust), but no
suitable emission was found. Either way, there is no strong emission line from this
interloper near the wavelength of HeII.

We detected an emission line to the SW of GN-z11 (bottom panel of Fig.~\ref{fig:interlopers}).
This source has no continuum and therefore no photometric redshift. We were thus unable to identify
its redshift. Clearly, however, no contamination can arise from this region over to the HeII clump.

Finally, we note that the HeII emission of the clump has a very marginal feature (less than 2$\sigma$) to the red side of the line (at $\sim 1.907~\mu m$). This does not match any other potential reasonably strong doublet or pair of lines that could be associated with a low-z interloper (e.g. H$\alpha$+[NII], [OII] doublet, [SII] doublet), not only in terms of wavelength separation but also because any reasonable pair of lines would imply the detection of a much stronger H$\alpha$ emission at other wavelengths. If the weak, marginal feature is confirmed, this could be an additional weaker companion at $\sim$690 km/s from the main system, or it could be tracing an outflow component. However, we refrain from speculating further on this marginal feature, as its significance is very low and requires additional observations.

\section{Spectrum extracted from large aperture excluding regions closer to GN-z11}\label{app:sp_ap_mod}

\begin{figure}%
\centering
\includegraphics[width=0.9\columnwidth]{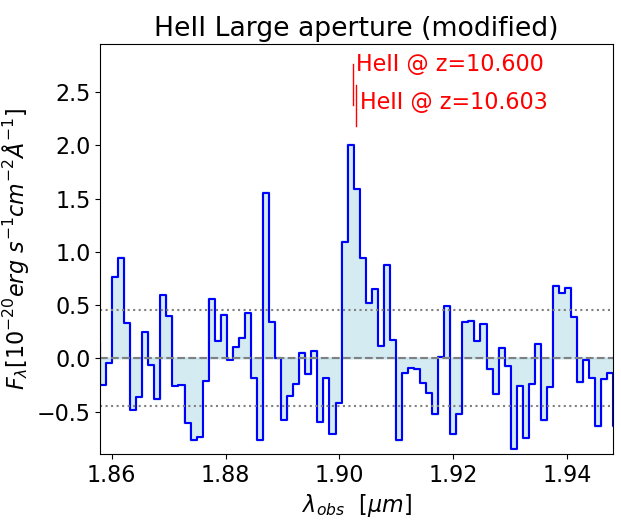}
\caption{Spectrum extracted from the same extracted aperture at the spectrum in the right panel of Fig.\ref{fig:spectra_ex}, that is, the aperture marked in yellow in Fig.\ref{fig:footprints} (Tab.\ref{tab:apertures}), but where the regions closer than 0.3$''$ to GN-z11 were excluded. The spectrum is essentially unchanged with respect to the spectrum extracted from the entire aperture.
}\label{fig:spheiimod}
\end{figure}

The HeII emission in GN-z11 is so faint that it cannot contribute to the spectrum shown in the right panel of Fig.\ref{fig:spectra_ex} extracted from the large aperture drawn with a yellow box in  Fig.\ref{fig:footprints} (Tab.\ref{tab:apertures}); otherwise, many other much stronger lines from GN-z11 would be seen in the spectrum. In order to further exclude any contamination, we re-extracted the spectrum from the same aperture by excluding the region closer than 0.3$''$ (i.e. six times the radius of the PSF at this wavelength) from GN-z11. The resulting spectrum is shown in Fig.\ref{fig:spheiimod}, which is essentially unchanged with respect to Fig.\ref{fig:spectra_ex}-right. If anything, the HeII detection is slightly more significant in this spectrum. Therefore, it is confirmed that the HeII flux from GN-z11 does not contribute to the flux observed in that aperture.

\section{Constraints on the HeII equivalent width}\label{app:ew}

An important parameter in this work is the determination of the EW of HeII.
In this Appendix, we provide some additional information on the methods to measure the EW.

As in most past spectroscopic studies at high redshift, inferring the EW directly from the spectra is extremely difficult, as the continuum is typically not detected in the spectrum. Therefore, most studies have resorted to inferring the continuum flux from photometry and combining it with the line flux inferred from the spectrum \citep[e.g.][]{Stark17,Debarros17,Mainali17,Jung20,Kusakabe20}. The availability of low-resolution spectroscopy with JWST has enabled detection of the continuum and measuring of the EW of lines directly in the spectra for a larger number of sources. However, with medium- and high-resolution spectroscopy, even with JWST, studies still have to rely on photometry to infer the EW of emission lines \citep[e.g.][]{Matthee23a}, and for faint sources, photometry is still used to determine the EW of lines, even with JWST low-resolution spectroscopic modes \citep[e.g.][]{Boyett22}.

In this study we are in a similar regime. In the IFU spectra, the continuum is totally undetected, and the source is so faint that the continuum is marginally constrained in the low-resolution MSA spectrum. Therefore, our primary method to determine the EW was by using the photometric information from the NIRCam imaging \citep[whose data processing is described in][]{tacchella_jades_2023}. Photometry was extracted from the same apertures used for the extraction of the IFS spectra, as indicated in Fig.\ref{fig:footprints}. We did not attempt any PSF correction when extracting the photometry, as both apertures are much larger than the PSF at the wavelength of the redshifted HeII, and thus the PSF does not really affect the photometry. Additionally, we were not dealing with a strong spiky source, which could potentially cause significant PSF losses. Finally, the PSF at the same wavelength is similar for NIRCam and for NIRSpec, so any residual second-order losses from the aperture should be the same with the two instruments. The photometry uncertainty was inferred as in \cite{tacchella_jades_2023}; specifically, we added the uncertainties from the error map in quadrature, which gives the quoted error. We compared this to the variance of fluxes measured in empty (only sky, no sources) apertures of the same size in the mosaic and found that they are comparable (within 10\%). Given that the continuum is marginally detected even in photometry, we adopted an approach similar to \cite{Matthee23a} for determining the EW of high-z galaxies with JWST spectroscopy.  \cite{Matthee23a} used the multi-band photometry to fit a Spectral Energy Distribution (SED) and then derived the continuum beneath the emission lines from the best-fit SED. In our case, the weak detections did not allow us to fit a proper SED. Moreover, using SED templates would imply making a priori assumptions on the nature of the continuum. We therefore adopted a simplified approach by inferring the continuum beneath the HeII with a linear interpolation of the photometry in the filters F150W, F200W, and F277W.
We also provided a solid lower limit on the EW by using the tentative continuum detection, as inferred above from photometry, and adding a 2$\sigma$ uncertainty, which gives the lower limit given in the main text of 20\AA.

Finally, we reported the 3$\sigma$ lower limit inferred from the non-detection of the continuum directly in the medium-resolution IFS spectrum.
We also attempted to estimate the EW directly from the prism spectrum, given that in this case the 
continuum is marginally detected in the spectrum (and it has been fitted with linear function, masking the HeII line itself). The resulting values and lower limits are given in Table.\ref{tab:emlines}

\section{NIRSpec-IFS vs NIRCam photometry comparison}\label{app:ifs-phot}

Given that some of our estimates of the EW were performed by measuring the continuum from the
photometry, it is important to verify the relative flux calibration of the
NIRSpec spectra with the NIRCam images.
The cross calibration between the MSA spectrum and photometry was already
checked by \cite{bunker_jades_2023} to be better than 5\% for the prism, so we do not
discuss it any further in this work.

Regarding the medium-resolution IFS spectra, we extracted the spectrum of
GN-z11 from the same aperture 0.6$''$ as used by \cite{tacchella_jades_2023} in the NIRCam
image. From the spectrum, we extracted photometry in the 
F200W filter, which is the closest to the HeII redshifted wavelength.
The aperture correction was applied,
but with such a large aperture and at this wavelength, the correction is
negligible. The resulting flux density is 141$\pm$17~nJy, which is consistent with the
NIRCam photometric measurement of 144$\pm$2.7~nJy. 
The small difference between the two fluxes in both filters is well below the uncertainties
resulting from the noise and background subtraction.
Similarly, in the F277W filter, the flux density inferred from the IFS spectrum is 106$\pm$13~nJy, compared to 121.7$\pm$4.2~nJy in the NIRCam image.

\end{document}